
\documentstyle{amsppt}
\magnification 1000
\NoBlackBoxes

\def\Z{\Bbb Z}
\def\C{\Bbb C}

\def\d{\partial}

\document

\centerline {\bf A unified representation-theoretic approach to
special functions}
\vskip .15 in
\centerline{\bf Pavel I. Etingof, Alexander A. Kirillov, Jr.}
\vskip .1in
\centerline{Department of Mathematics, Yale University, New Haven, CT
06520, USA}
\centerline{e-mail etingof\@math.yale.edu,
			kirillov\@math.yale.edu}
\vskip .1in
\centerline{\bf Dedicated to I.M.Gel'fand on the
occasion of his  80th birthday}

A representation-theoretic approach to special
functions was developed in the 40-s and 50-s in the works
of I.M.Gelfand, M.A.Naimark, N.Ya.Vilenkin, and their collaborators
(see \cite{V},\cite{VK}). The essence of this approach is the fact
that most classical special functions can be obtained as suitable
specializations of matrix elements or characters
of representations of groups.
Another rich source of
special functions is the theory of Clebsch-Gordan
coefficients which describes the geometric juxtaposition of
irreducible components inside the tensor product of two
representations (cf. \cite{VK}).
Finally, in recent works on representations of (quantum)
affine Lie algebras it was shown that matrix elements of
intertwining operators between certain representations of these
algebras are interesting special functions --
(q-)hypergeometric functions and their
generalizations \cite{TK, FR}.

In this paper we suggest a general method of getting special functions
from representation theory which unifies the three methods mentioned
above and allows one to define and study many new special functions.
We illustrate this method by a number of examples --
Macdonald's polynomials, eigenfunctions of the
Sutherland operator, Lam\'e functions. Other examples
will be described in our future papers.

{\bf 1. Vector-valued spherical functions. }
Let $\Cal H$ be a Hopf algebra over $\C$,
$H\subset\Cal H$ be a subgroup of group-like elements,
$V,W,U$ be irreducible $\Cal H$-modules. Let $\Phi:V\to W\otimes U$
be an intertwining operator for $\Cal H$. From this data one can
construct two kinds of special functions.

{\it 1. Vector-valued matrix elements. } Set
$f_{vw\Phi}(h)=<w,\Phi hv>\in U$, $h\in H$, $v\in V$,$w\in W^*$.
We call this function on $H$ a $U$-valued matrix
element.

{\it 2. Vector-valued characters. } Let $V=W$.
Set $\chi_{\Phi}(h)=\text{Tr}\mid _{V}(\Phi h)\in U$, $h\in H$.
We call this function on $H$ a $U$-valued character.

{\bf Example 1. } If $U=\C$, $V=W$, $\Phi=\text{Id}$, and
we get the classical
matrix elements and characters.

{\bf Example 2. } The numbers $<f_{vw\Phi}(e),u>$, $u\in U^*$,
where $e$ is the identity element in $H$,
are the Clebsch-Gordan coefficients for $\Cal H$
(we assume that $v,w,u$ are taken from some bases of $V,W^*,U^*$).

{\bf Example 3. } Let $\Cal H$ be the quantum affine algebra
$U_q(\hat{\frak g})$ corresponding to a finite -dimensional simple Lie
algebra $\frak g$ ($q$ can be equal to $1$), $H=\{e\}$. Let
$V,W$ be Verma modules over $\Cal H$ with the same central charge,
and let $U=U_1(z_1)\otimes \dots\otimes U_n(z_n)$, where $z_i\in
\C^*$, $U_i$ are finite-dimensional representations of $\Cal H$,
and $U_i(z)$ denotes the representation of $\Cal H$ in the space $U_i$
defined by $\pi_{U_i(z)}(a)=z^{\text{deg}(a)}\pi_{U_i}(a)$, $a\in \Cal
H$, $\text{deg}(a)$ being the homogeneous degree of $a$.
Let $\Phi: V\to W\otimes U$
be a vertex (=intertwining) operator
(here $\otimes$
should be understood as a completed tensor product).
Let $v,w$ be the vacuum vectors of $V,W^*$.
(For more details see \cite{FR}).
Then the $U_1\otimes\dots\otimes U_n$-valued matrix element
$f_{vw\Phi}(e)$ is a function of $n$ complex variables $z_1,...,z_n$.
For $q=1$ such functions arise in conformal field theory as
correlation functions on the sphere; for $q\ne 1$, they appear as
correlation functions of solvable lattice models in statistical
mechanics.
It is known that  they satisfy the (classical or quantum)
Knizhnik-Zamolodchikov (KZ) equations
and express via hypergeometric (q-hypergeometric) functions and their
generalizations. Thus, solutions of (quantum) KZ
equations can be regarded as Clebsch-Gordan coefficients for
(quantum) affine Lie algebras.

{\bf 2. A generalized Peter-Weyl theorem. }
In this section $H=G$ is a finite group or a compact Lie group, and
$\Cal H$ is the convolution algebra of (generalized) functions on $G$.
Notation: $dg$ is the invariant probability measure on $G$;
$R(G)$ is the set of irreducible unitary representations of $G$;
$U\in R(G)$; $(\cdot,\cdot)_U$ is the inner product in
$U$; $L^2(G,U)$ is the Hilbert space of $U$-valued $L^2$-functions on
$G$ with respect to $dg$, with the inner product $<f_1,f_2>=
\int_G(f_1(g),f_2(g))_Udg$; $L^2(G,U)^G$ is the subspace of
$L^2(G,U)$ spanned by all functions $\chi$ such that
$\chi(hgh^{-1})=h\chi(g)$, $g,h\in G$.

For every $V\in R(G)$, fix an orthonormal basis $B_V$ of $V$.
For any $V,W\in R(G)$, let $X_{VW}$ be
the space of $G$-homomorphisms: $\Phi: V\to W\otimes U$.
This is a Hilbert space since it is isomorphic to the space of
$G$-invariants in $V^*\otimes W\otimes U$. Let $\Cal B_{VW}$ be an
orthonormal basis in $X_{VW}$. Let $f_{vw\Phi}\in L^2(G,U)$ be defined by
$f_{vw\Phi}(g)=(w,\Phi g v)_W\in U$, $g\in G,v\in V,w\in W,\Phi\in X_{VW}$.
We have $f_{vw\Phi}(gh)=f_{(hv)w\Phi}(g)=gf_{v(g^{-1}w)\Phi}(h)$.
Also, define the functions $\chi_{\Phi}(g)=\text{Tr}\mid_V(\Phi g)=
\sum_{v\in B_V}f_{vv\Phi}(g)$, $\Phi\in X_{VV}$. Clearly, $\chi_{\Phi}\in
L^2(G,U)^G$.

{\bf Theorem. } \sl (i) $\{(\text{dim}V\text{dim}W)^{1/2}
f_{vw\Phi}:v\in B_V,w\in B_W,\Phi \in \Cal B_{VW},
V,W\in R(G)\}$ is an orthonormal basis of $L^2(G,U)$.

(ii) $\{\chi_{\Phi}:\Phi\in \Cal B_{VV}, V\in R(G)\}$ is an orthonormal basis
of $L^2(G,U)^G$.\rm

When $U=\C$, this theorem reduces to the standard
Peter-Weyl theorem \cite{K}.

{\bf Remark. } This theorem can be generalized to the case of quantum
groups (cf. \cite{EK1}).

{\bf 3. Macdonald's polynomials as vector-valued characters.}
Let $G=SU(n)$, $\Cal H$ be the algebra of functions on $G$, and let
$H$ be the Cartan subgroup of $G$ (diagonal matrices).
Macdonald's polynomials $P_{\lambda}(q,k,h)$ ($k\in \Bbb Z$, $h\in H$,
$\lambda$
is a dominant integral weight for $G$) are uniquely defined
by the following properties:

1. $P_{\lambda}(q,k,\cdot)$ are trigonometric polynomials on $H$
symmetric under the action of the Weyl group $S_n$.

2. For fixed $q$ and $k$,
$P_{\lambda}(q,k,\cdot)$ are orthogonal on $H$ with respect to
the weight $|\Delta|^2$, where $\Delta(q,k,h)=
\prod_{m=0}^{k-1}\prod_{\alpha>0}
(1-q^me^{<\alpha,\xi>})$, where $\xi\in \frak h=\text{Lie}H$ is such that
$h=e^{\xi}$, and $\alpha$ runs over all positive roots of $G$.

3. $P_{\lambda}=\chi_{\lambda}+\sum_{\nu<\lambda}c_{\lambda\nu}
\chi_{\nu}$, where $\nu$ is a dominant integral weight of $G$, and
$\chi_{\nu}$ is the character of the irreducible representation of $G$
with highest weight $\nu$, and $c_{\lambda\nu}$ are constants
depending on $q,k$.

{\it Example: } $P_{\lambda}(q,1)=\chi_{\lambda}$.

In the special case $q=1$, Macdonald's polynomials are
called Jack's symmetric functions.

Let $V=L_{\lambda}$ be the finite-dimensional irreducible
representation of $G$ with highest weight $\lambda$. Let
$U=S^{kn}\C^n$, where $k$ is a positive integer, be a representation
of $G$. Let $\Phi$ be a nonzero intertwining operator $V\to V\otimes U$.
Such an operator exists iff $\lambda\ge k\rho$, where $\rho$ is the
half-sum of positive roots of $G$, and if it exists, it is unique
up to a factor.
Let $W(h)$ be the Weyl denominator:
$W(h)=\prod_{\alpha>0}(e^{\frac{1}{2}<\alpha,\xi>}-
e^{-\frac{1}{2}<\alpha,\xi>})$, $h=e^{\xi}$.
Let $\lambda=\nu+k\rho$, where $\nu$ is any dominant integral weight.
Define the functions $\psi_{\nu}(k,h)=
W(h)^{-k}\text{Tr}|_V(\Phi
h)$. These functions take values in the zero weight component of $U$,
which is one-dimensional.
Thus, we can regard them as scalar functions, choosing the normalization
in such a way that the coefficient to $e^{<\nu,\xi>}$ in $\psi_{\nu}$
is 1.

{\bf Theorem. }\cite{EK1} \sl The functions $\psi_{\nu}(k,h)$ are
the Jack's symmetric functions $P_{\nu}(1,k+1,h)$.\rm

{\it Remark. } The orthogonality of $\{\psi_{\nu}\}$ immediately follows
from the generalized Peter-Weyl theorem.

In the special case $G=SU(2)$ the polynomials $P_{\nu}(1,k+1,h)$
are the classical Gegenbauer polynomials --
the even trigonometric polynomials in one variable orthogonal with respect
to the weight $\text{sin}^{2k+2}x$.

Let now $\Cal H$ be the quantum group $U_q(\frak{sl}_n)$.
The rest of notation is as above (the modules
$L_{\lambda}$ and $S^{kn}\C^n$ are the $\Cal H$-modules
obtained by q-deformation of the corresponding $SU(n)$-modules).
Let $W_{q,m}(h)=\prod_{m=0}^{k-1}\prod_{\alpha>0}
(q^me^{\frac{1}{2}<\alpha,\xi>}-
q^{-m}e^{-\frac{1}{2}<\alpha,\xi>})$, $h=e^{\xi}$.
Define the functions
$\psi_{\nu}(q,k,h)=W_{q,k}(h)^{-1}\text{Tr}|_V(\Phi
h)$, with the normalization described above.

{\bf Theorem.}\cite{EK1} \sl The functions $\psi_{\nu}(k,h)$ are
the Macdonald's polynomials $P_{\nu}(q^2,k+1,h)$.\rm

{\bf 4. Diagonalization of the Sutherland operator.}
The Sutherland operator is the Hamiltonian of a quantum
many-body problem (see \cite{OP}):
$H=-\frac{1}{2}\sum_{j=1}^n\frac{\d^2}{\d x_j^2}
+C\sum_{i<j}\text{sinh}(x_i-x_j)^{-2}$ ($C$ is a constant).
It is known \cite{OP} that $H$ has $n$ functionally independent
commuting quantum integrals $L_1,L_2,...,L_n$:
$L_m=\sum_{j=1}^n\frac{\d^m}{\d x_m}+\text{lower order terms}$
($L_1=\sum_{j=1}^n\frac{\d}{\d x_j}$, $L_2=-2H$), and $[L_i,H]=0$ for
all $i$. Therefore, it makes sense to consider
the system of differential equations $L_i\psi=\Lambda_i\psi$,
$\Lambda_i\in\C$ (the eigenvalue problem). For each set of eigenvalues
$\Lambda_i$ this system has $n!$ linearly independent solutions.
For special values of $\{\Lambda_i\}$ one of these solutions
expresses via Jack's polynomials, but in general solutions are
transcendental and express via generalized hypergeometric functions.
Here we interpret these solutions as vector-valued characters
for $\frak{gl}_n$.

Let $\Cal H=U(\frak{gl}_n)$, $\frak h=\C^n$ be the Cartan subalgebra
in $\frak{gl}_n$, $H=\text{exp}(\frak h)$.
Let $V$ be the Verma module over $\frak{gl}_n$ with highest weight
$\lambda\in\C^n$. Let $\mu\in\C$ be such that $C=\mu(\mu+1)$,
and let $W$ be the module
spanned by all functions of the form $z_1^{\mu+m_1}z_2^{\mu+m_2}\dots
z_n^{\mu+m_n}$, $m_j\in \Z$, $\sum_jm_j=0$, with the action of
$\frak{gl}_n$ given by $E_{ij}\mapsto z_i\frac{\d}{\d z_j}-\mu\delta_{ij}$
($(E_{ij})_{kl}=\delta_{ik}\delta_{jl}$).
Let $\Phi:V\to V\otimes W$ be an intertwining operator for
$\frak{gl}_n$ ($\otimes$ is the completed tensor product).
$\Phi$ is unique up to a factor if $V$ is irreducible (which
happens for a generic $\lambda$).

Define the functions
$\phi_{\lambda}(x_1,...,x_n)=W(e^{\xi})^{-1}\text{Tr}|_V(\Phi
e^{\xi})$, where $\xi=2\sum_{j=1}^n x_jE_{jj}$ ($W$ is the Weyl denominator).
As in Section 3, we regard $\phi_{\lambda}$ as scalar functions.

{\bf Theorem. }\cite{E} \sl (i) The function $\psi_{\lambda}$ satisfies
the system of equations $L_i\psi_{\lambda}=\Lambda_i\psi_{\lambda}$,
where $\Lambda_i=p_i(\lambda+\rho)$, and $p_m$ are symmetric
polynomials with the highest term proportional to $\sum_{j}\lambda_j^m$.

(ii) For a generic $\lambda$, the functions
$\psi_{\sigma(\lambda+\rho)-\rho}$, $\sigma\in S_n$, form a basis
in the space of solutions of the system
$L_i\psi=p_i(\lambda+\rho)\psi$.\rm

{\it Remark. } This theorem can be generalized to the case of quantum
$\frak{gl}_n$, which provides a diagonal basis
of functions for Macdonald's difference operators (cf. \cite{Ch})

{\bf 5. Lam\'e functions as vector-valued characters
of $\widehat{sl}_2$. }
We preserve the notations of Example 3 from Section 1. Let $\frak g=
\frak{sl}_2$, $V=W=M_{\lambda,k}$ is the Verma module with highest
weight $\lambda$ and central charge $k$, $n=1$, $z_1=1$,
and $U_1$ is the same as in Section 4. Consider the function
$$
F(x,\tau)=\left(\text{Tr}|_{M_{k/2,k}}(e^{-\pi i(2\tau
d+(x+\frac{\tau}{2})h)})\right) ^{-1}
\text{Tr}|_V(\Phi e^{-\pi i(2\tau d+(x+\frac{\tau}{2})h)}),
$$
where $d$ is the homogeneous gradation operator and $h=\text{
diag}(1,-1) \in\frak{sl}_2$. This is a 1-point correlation
function of conformal field theory on the torus.
As in Sections 3,4, we regard $F$ as a
scalar function.

{\bf Theorem. }\cite{EK} \sl $F$ satisfies the Schr\"odinger
equation $-2\pi i(k+2)\frac{\d F}{\d\tau}+\frac{\d^2 F}{\d x^2}=
\mu(\mu+1)(\wp(x+\frac{\tau}{2},\tau)+c)F$, where $\wp$ is the Weierstrass
elliptic $\wp$-function, and $c$ is a constant. \rm

If $\mu$ is a positive integer, it is possible to express $F$ as
a $\mu$-dimensional integral of products of powers of theta-functions
and exponents. When $k\to -2$ (critical level), the above equation
becomes Lam\'e equation \cite{WW};
finding the asymptotics of the integrals,  we
recover classical formulas from \cite{WW}, expressing  Lam\'e functions via
 theta-functions (see \cite{EK}).

{\bf Acknowledgements.} We would like to thank our advisor Professor
Igor Frenkel for stimulating discussions and Professors I.Cherednik, R.Howe, D.
Kazhdan and A.Varchenko for  helpful remarks.

\end